\documentclass[aps,prd,nofootinbib,reprint]{revtex4-1}
\usepackage[utf8]{inputenc}
\usepackage{amsmath}
\usepackage{amsfonts}
\usepackage{amssymb}
\usepackage{color}
\begin{document}
\title{Rotating metric in Non-Singular Infinite Derivative Theories of Gravity}
\author{Alan S. Cornell,$^{1}$~Gerhard Harmsen,$^{1}$~Gaetano Lambiase,$^{2,~3}$~Anupam Mazumdar,$^{4,~5}$}
\affiliation{$^{1}$~National Institute for Theoretical Physics; School of Physics, University of the Witwatersrand, Johannesburg, Wits 2050, South Africa.}
\affiliation{$^{2}$~Dipartimento di Fisica ``E.R Caianiello'', Universit\`{a} di Salerno, I-84084 Fisciano (SA), Italy.}
\affiliation{$^{3}$~INFN - Sezione di Napoli, Gruppo collegato di Salerno, I-84084 Fisciano (SA), Italy.}
\affiliation{$^{4}$~Van Swinderen Institute, University of Groningen, 9747 AG, Groningen, The Netherlands.}
\affiliation{$^{5}$~Kapteyn Astronomical Institute, University of Groningen, 9700 AV Groningen, The Netherlands.}
\date{\today}
\begin{abstract}
In this paper we will provide a non-singular rotating space time metric for a ghost free infinite derivative theory of gravity in a linearized limit. We will provide
the predictions for the Lense-Thirring effect for a slowly rotating system, and how it is compared with that from general relativity.
\end{abstract}
 \vskip -1.0 truecm

\maketitle

\section{Introduction}

Einstein's theory of General Relativity (GR) has been an extremely successful theory of gravity in predicting numerous observational tests at large distances and late time scales, i.e. the infrared (IR), matching observations ranging from solar system tests to large scale structures of the Universe~\cite{Will:2014kxa} . The recent discovery of gravitational waves from the binary blackhole mergers have added yet another interesting dimension towards the success story of GR~\cite{Abbott:2016blz}. In spite of all these successes, GR has problems at short distances and small time scales, i.e. the ultraviolet (UV) regime, where  blackhole and cosmological singularities are inevitable.

Beside classical pathologies, the 2-derivative action of  GR poses problems at a quantum level. Pure GR is one-loop renormalizable~\cite{Hooft}.
Also, the quadratic curvature gravity with 4-derivatives is also a power-counting renormalizable theory of gravity~\cite{Stelle:1976gc}. However, there are
problems due to the presence of a massive {\it ghost} in the spin-2 component for the quadratic curvature gravity, which leads to an unstable vacuum. This is reminiscent of any higher derivative classical theory, irrespective of the spin, where the Hamiltonian density becomes unbounded from below, known as Ostr\"ogradsky  instability~\cite{Ostrogradsky}. Therefore, any modification of GR will always lead to extra propagating degrees of freedom, which are required to be tamed in order to make sure that these degrees of freedom are not {\it ghost}-like.

Recently, it has been shown that an {\it infinite derivative theory} of covariant gravity  (IDG) can be made {\it ghost free} and also singularity free~\cite{Biswas:2011ar,Biswas:2005qr}. Such actions have also been motivated from string theory, see~\cite{Tseytlin:1995uq,Siegel:2003vt}.
The gravitational potential for a {\it static} and {\it spherically symmetric} metric asymptotes to a $1/r$-law at large distances in the IR~\cite{Edholm:2016hbt}, but in the UV the potential becomes {\it constant}, and the gravitational
force $F_g\rightarrow 0$, signaling {\it classical asymptotic freedom} at short distances for the gravitational interaction within the IDG. The solution provides a non-singular compact object and the gravitational potential remains linear throughout the region of space time, with $mM\leq M_p^2$~\cite{Biswas:2011ar}, where $m$ is the mass of the source, $M_p=1/\sqrt{8\pi G}=2.4\times 10^{18}$~GeV, and $M$ denotes the scale of non-locality, which plays an important role during graviton interactions. If one preserves the {\it area}-law of gravitational entropy then the scale of non-locality also {\it shifts} from $M$ to $M_{eff}\sim M/\sqrt{N}$, where $N$ defines the number of states involved in the collapse process. The shifting in the scale of non-locality can potentially resolve the singularity and the horizon problem for massive compact objects (with mass much above the solar mass)~\cite{Koshelev:2017bxd}. The dynamical solution for ghost free and singularity free IDG has also been investigated in the regime where the metric potentials are bounded below unity, and it was found that no trapped surface is formed and that no curvature singularity is ever developed in the non-rotating case~\cite{Frolov:2015bia,Frolov:2015bta,Frolov:2016pav}. Furthermore, at a quantum level, such a class of theories {\it also} hints to a UV finiteness of gravitational interactions~\cite{-Yu.-V.,Tomboulis,Modesto,Talaganis:2014ida}, i.e. beyond 1-loop the theory becomes UV finite.

The aim of the current paper is to seek a rotating solution for a ghost free IDG, i.e. equivalent of a Kerr-metric~\cite{Kerr:1963ud}, within a linearized limit. We will first use the case of a {\it slowly rotating } source, and then match the results with the one obtained from the static solution to the rotating case by employing the Demanski-Janis-Newmann (DJN) algorithm~\cite{Demianski:1972uza}, which is able to obtain a solution for the rotating case, without solving the Einstein equations directly~\cite{Erbin:2014aja}. The method works by taking a static spherically symmetric metric, converting to a null metric, and then complexifying the radial and null time coordinates of the metric. There are no strict rules on how these transformations must be completed, but we must ensure that the new functions created after the transformations are real. Once we have the new complexified coordinates, and new radially dependent terms, we must then perform our transformation to the null rotating coordinates and finally to the Boyer-Linquist coordinates, for a detailed discussion, see \cite{Erbin:2014aja}. The original method, as developed by Newman and Janis, requires the use of the Newman-Penrose tetrad formalism in finding a null tetrad basis \cite{Newman:1965my}. However, in Ref.~\cite{Giampieri} the author has updated the method without using Newman-Penrose tetrads, and therefore making the method easier to follow and allowing us to tackle more complex solutions beyond those found in GR~\cite{Erbin:2014aja}.

\section{Infinite Derivative Gravity and slowly rotating metric}

Let us now proceed with the most general covariant, quadratic curvature, torsion free, IDG action, which has been derived around constant curvature backgrounds, see~\cite{Biswas:2011ar,Biswas:2013kla,Biswas:2016egy}:
\begin{eqnarray}\label{action}
S = \frac{1}{2\kappa^2} \int d^{4}x\sqrt{-g} \left( R + \alpha\left[ R{\cal F}_1(\Box_{s})R\right.\right. \\ 
\left.\left.+ R_{\mu\nu}{\cal F}_2(\Box_{s})R^{\mu\nu} +R_{\mu\nu\alpha\beta}{\cal F}_3(\Box_{s})R^{\mu\nu\alpha\beta}\right] \right)\,, \nonumber
\end{eqnarray}
where $\kappa^2=8\pi G$, and $\alpha$ is a dimensionfull coupling. The indices run from; $\mu,~\nu =0,1,2,3$, and we mostly use the $(-,+,+,+)$ signature.
The three form factors, $\mathcal{F}_{i}(\Box_s)=\sum_{n=0}^{\infty}c_{i n}\Box_s^{n}$,
are functions of the infinite order covariant differential operators, $\Box_{s}\equiv \Box/M^{2}$, and the infinite coefficients $c_{in}$ are fixed by
demanding that the above action contains only the massless transverse traceless graviton degrees of freedom, i.e. massless spin-2 and spin-0 components. Around the Minkowski spacetime this constrains the form factors to be; $2{\cal F}_1+{\cal F}_2+2{\cal F}_3=0$~\cite{Biswas:2011ar,Biswas:2013kla,Buoninfante:2016iuf}. 

The d'Alembert operator is denoted as; $\Box=g^{\mu\nu}\nabla_{\mu}\nabla_{\nu}$, and $M$ is the new scale of physics, which signifies the
scale of non-locality in this context, i.e. the interactions of the above theory becomes non-local beyond $M$. In the limit when $M\rightarrow \infty$,
the above action reduces to a pure Einstein-Hilbert action, with the gravitational potential in the IR behaving as $\sim 1/r$, at large distances. The
best constraint on $M$ arises from observing the departure from the Newtonian potential, which has not been observed beyond $5\times 10^{-6}$m~\cite{Kapner}, placing the constraint on $M\geq 0.004$~eV~\cite{Edholm:2016hbt}.

Let us consider the equations of motion for the above action, Eq.~(\ref{action}) in the linearized limit~\cite{Biswas:2013cha}, i.e. $g_{\mu\nu}=\eta_{\mu\nu}+h_{\mu\nu}$, by neglecting the higher order
terms in the perturbation, ${\cal O}(h_{\mu\nu}^2)$, and imposing the De Donder gauge $\partial^{\mu} (h_{\mu\nu}-(1/2) \eta_{\mu\nu}h)=0$.
Solving the equations of motion for the metric:
\begin{equation}\label{ThisOne}
\begin{aligned}
ds^{2}=-\left(1+2\Phi \right)dt^{2} + 2\vec{\textbf{h}}\cdot d\textbf{x}dt + \left(1-2\Phi \right)d\textbf{x}^{2}\,,
\end{aligned}
\end{equation}
with $T_{00}=\rho$ and $T_{0i}=-\rho v_{i}$, where $\rho=m\delta^3(\vec{r})$, $m$ is the mass of the source and $v_{i}$ is the velocity of the source,
we get:
\begin{equation}
\begin{aligned}\label{eqn}
4a(\Box_{s})\Box\Phi=2\kappa\rho\,,~~~~~~
a\left(\Box_{s}\right)\Box h_{0i}=-\kappa\rho v_{i}\, ,\\
\end{aligned}
\end{equation}
where the coefficient $a(\Box_s)$ is determined by demanding that the gravity remains massless and does not introduce any new dynamical degrees of freedom. The function $a(\Box_ s)$ should be an {\it exponential of an entire function}~
\cite{Biswas:2011ar,Biswas:2005qr,Biswas:2013kla}, where one simple choice is:
\begin{equation}\label{aBox}
\begin{aligned}
a\left(\Box_{s}\right)=e^{-\Box/M^{2}}\,.
\end{aligned}
\end{equation}
We have selected the velocity of the source to be such that the angular velocity points along the z axis, this allows us to define the velocities as
\begin{equation}
\begin{aligned}
v_{x}=-y\omega,~~~~ v_{y}=x\omega\,.
\end{aligned}
\end{equation}
This is the case of a very slowly rotating object, which would experience very little flattening of the metric.
By taking a Fourier transform of the components in Eq.(\ref{eqn}), we obtain:
\begin{equation}\label{PhiSol}
\begin{aligned}
\Phi(r)\approx & \frac{2m}{8\pi^{2}M_{p}^{2}r}\int\frac{dp}{p}e^{-\left(\frac{p}{M}\right)^{2}}\sin(pr)\\
= &\frac{m}{8\pi M_p^2 r}\textbf{Erf}\left(\frac{rM}{2} \right)\,,
\end{aligned}
\end{equation}
where $M_p^2 =1/(8\pi G)$. Note that as $r\rightarrow\infty$, the error function $\rightarrow 1$, and we recover the GR limit, while as $r\leq 2/M$ the error function goes
linearly with the argument, such that the potential $\Phi\sim mM/M_{p}^{2}$, becomes constant, see also~\cite{Buoninfante:2018xiw}. The current bound on $M$ arises precisely from the
IR limit, see~\cite{Edholm:2016hbt}.
The solution for  $h_{0i}$ can be solved analogously, $a(\Box_s)\Box h_{0x}=y\kappa\omega\rho$,~~$a(\Box_s)\Box h_{0y}=-x\kappa\omega\rho$,
where $\omega=\textbf{v}/r$ and $\textbf{v}$ is a constant velocity,
and $a(\Box_s)\Box h_{0z}=0$. We obtain:
\begin{equation}\label{OffDiagCart}
\begin{aligned}
h_{0x}=4y\omega\Phi,~~~~~
h_{0y}=-4x\omega\Phi,~~~~h_{0z}=0\,,
\end{aligned}
\end{equation}
and the resulting metric is given by:
\begin{equation}\label{MetricCart}
\begin{aligned}
ds^{2}=& -\left(1+2\Phi \right)dt^{2} +4y\omega\Phi dt dx-4x\omega\Phi dtdy \\
&+\left(1-2\Phi \right)d\text{x}^{2}\,.\\
\end{aligned}
\end{equation}
Furthermore, by using the standard conversion from Cartesian to radial coordinates,  we obtain the metric in Boyer-Lindquist coordinates:
\begin{equation}\label{IDGMetGae}
\begin{aligned}
ds^{2}=& -\left(1+2\Phi \right)dt^{2} -4\omega r^{2}\sin^{2}\theta\Phi d\phi dt\\
& +\left(1-2\Phi \right)\Big( dr^{2} +r^{2} d\theta^{2} + r^{2}\sin^{2}\theta d\phi^{2}\Big)\,.\\
\end{aligned}
\end{equation}
We will define the angular momentum, $J$, as
$\textbf{v}=(r\times J)/(m r^{2})$, and rewrite the metric for the rotating source in the form:
\begin{equation} \label{IDGMet}
\begin{aligned}
ds^{2}=& -\left(1+2\Phi \right)dt^{2} -4\frac{J\sin^{2}\theta}{m}\Phi d\phi dt\\
& +\left(1-2\Phi \right)\Big( dr^{2} +r^{2} d\theta^{2} + r^{2}\sin^{2}\theta d\phi^{2}\Big)\,.\\
\end{aligned}
\end{equation}
The above metric is a very good example of a slowly rotating object, which can be used to probe the deviation from a rotating metric in GR and IDG.

\section{ Demanski-Janis-Newmann algorithm}

Now, we will show that the rotating metric can also be obtained for a non-singular static metric by employing the DJN algorithm.
Our starting point will be the following static metric:
\begin{equation}
\begin{aligned}
ds^{2}=-\left(1+2\Phi \right)dt^{2} + \left(1-2\Phi \right)d\text{\bf x}^{2},
\end{aligned}
\end{equation}
where $\Phi$ is exactly the same as in Eq.~(\ref{PhiSol}). This metric can be rewritten in a spherical coordinate system as follows
\begin{equation}
\begin{aligned}
ds^{2}=-f_{t}dt^{2} + f_{r}dr^{2} + f_{\Omega}\left(d\theta^{2} + \sin^{2}\theta d\phi^{2} \right),
\end{aligned}
\end{equation}
where $f_{t}=1+2\Phi$, $f_{r}=1-2\Phi$ and $f_{\Omega}=r^{2}f_{r}$. In the Eddington-Finkelstein coordinates the metric will read:
\begin{equation}
\begin{aligned}
ds^{2}=-f_{t}du^{2}-2\sqrt{f_{t}f_{r}}dudr+f_{\Omega}\left(d\theta^{2}+\sin^{2}\theta d\phi^{2} \right)\,,
\end{aligned}
\end{equation}
where  $~t=u+(f_{r}/f_{t})r$ has been used. As discussed in Refs. \cite{Erbin:2014aja, Demianski:1972uza}, we must now complexify the coordinates
$u$ and $r$, as follows:
\begin{equation}
\begin{aligned}
r\rightarrow r'=r+ai\cos\theta\,,~~~~~u\rightarrow u'=u-ai\cos\theta\,,
\end{aligned}
\end{equation}
where $a$ is a rotation parameter, and is related to the angular momentum:
\begin{equation}
a= {J}/{m}\,.
\end{equation}
Using the ansatz $id\theta=\sin\theta d\phi$~\cite{Giampieri,Erbin:2014aya,Ferraro:2013oua}, our differentials transform as $dr=dr'-a\sin^{2}\theta d\phi$ and $du=du'+a\sin^{2}\theta d\phi$. In the DJN approach we must choose a transformation for $r$, $r^{2}$ and $1/r$, where
these transformations must ensure that the functions $f_{i}$  remain real and the $\theta$ dependence is purely
$\cos\theta$, such that:
\begin{equation}
\begin{aligned}
r\rightarrow r\,,~~~\frac{1}{r}\rightarrow\frac{\text{Re}(r')}{|r'|^{2}}\,,~~~r^{2}=|r'|^{2}.
\end{aligned}
\end{equation}
Therefore, our functions transform as
\begin{equation}
\begin{aligned}
f(r)\rightarrow\widetilde{f}_{t,r}(r,\theta) &= 1 \pm 2\frac{m r}{8\pi M_{p}^{2}\Sigma}\textbf{Erf}\left(\frac{rM}{2} \right),\\
r^{2}\rightarrow &\Sigma \equiv r^{2}+a^{2}\cos^{2}\theta.\\
\end{aligned}
\end{equation}
Using these transformations,  we obtain the following null rotating metric \cite{Erbin:2014aja}:
\begin{equation}\label{NullRotate}
\begin{aligned}
ds^{2} =& - \widetilde{f}_{t}\left(du + \alpha dr+ \omega \sin\theta d\phi \right)^{2} + 2\beta drd\phi \\
&+\Sigma\widetilde{f}_{r}\left(d\theta^{2}+\sigma^{2}\sin^{2}\theta\phi^{2} \right),
\end{aligned}
\end{equation}
where
\begin{equation}
\begin{aligned}
\omega = a\sin\theta - &\sqrt{\frac{\widetilde{f}_{r}}{\widetilde{f}_{t}}}a\sin\theta\,,~~~\sigma^{2} = 1 + \frac{a^{2}\sin^{2}\theta}{r^{2}+a^{2}},\\
&\alpha = \sqrt{\frac{\widetilde{f}_{r}}{\widetilde{f}_{t}}}\,,~~~\beta =-\widetilde{f}_{r}a\sin^{2}\theta.
\end{aligned}
\end{equation}
In order to convert this null metric into the Boyer-Lindquist form, we must ensure that the functions
\begin{equation}
\begin{aligned}
 g(r)=\frac{\sqrt{\left(\widetilde{f}_{t}\widetilde{f}_{r}\right)^{-1}}\widetilde{f}_{\Omega}-F'G'}{\Delta}\,,~~
 h(r)=\frac{F'}{H(\theta)\Delta}
\end{aligned}
\end{equation}
are functions of $r$ only, where $\Delta = (\widetilde{f}_{\Omega}/\widetilde{f}_{r})\sigma^{2}$.
This is trivially true for $h(r)$, but is {\it only} true for $g(r)$ if $\Phi\ll1$, such that $f^{-1}_{r}=f_{t}$. In the above metric we are considering small perturbations on the Minkowski metric and so this statement is true, and we are therefore allowed to perform the transformation. We use the solution as given in \cite{Erbin:2014aja} and after some algebra we obtain:
\begin{equation}\label{DNJMetric}
\begin{aligned}
ds^{2}=&-(1+2\widetilde{\Phi})dt^{2} - 4a\widetilde{\Phi}\sin^{2}\theta d\phi dt +\frac{\Sigma (1-2\widetilde{\Phi})}{r^{2}+a^{2}}dr^{2}\\
& +\Sigma(1-2\widetilde{\Phi})\left( d\theta^{2} + \sin^{2}\theta\left(\frac{r^{2}+a^{2}}{\Sigma} \right)d\phi^{2} \right)\,,
\end{aligned}
\end{equation}
where
\begin{equation}
\widetilde{\Phi}=\frac{mr}{8\pi M_{p}^{2}\Sigma}\textbf{Erf}\left(\frac{rM}{2} \right).
\end{equation}
Note that for a slowly rotating case we recover Eq.(\ref{IDGMet}), in this case, $r^{2}+a^{2}\cos^{2}\theta\approx r^{2}$, since $a\rightarrow 0$,
which implies that $\widetilde{\Phi}\rightarrow\Phi$, hence Eq. (\ref{DNJMetric}) can be rewritten as
\begin{equation}\label{DNJ-1}
\begin{aligned}
ds^{2}=&-(1+2\Phi)dt^{2} - 4\frac{J\Phi\sin^{2}\theta}{m} d\phi dt\\ &+(1-2\Phi)\left(dr^{2} +r^{2}d\theta^{2} + r^{2}\sin^{2}\theta d\phi^{2} \right)\,.
\end{aligned}
\end{equation}
Indeed the two metrics, see Eq.~(\ref{IDGMet}) and Eq.~(\ref{DNJ-1}), are identical in this limit.

The IR limit of Eq. (\ref{DNJMetric}), when $r\rightarrow \infty$, we obtain the potential $\widetilde{\Phi}\rightarrow 1/r$, which is similar to that of the GR limit. In the case of $r\leq 2/M$, the potential reduces to $\widetilde{\Phi}\rightarrow (mM r^{2})/(16\pi M_{p}^{2}\Sigma)$. If $a< r < 2/M$, then the metric potential reduces to that of the static limit, i.e. $\widetilde \Phi \sim \Phi \sim mM/M_p^2$.

On the other hand, if the angular momentum is large, i.e. $a\gg r$, then the metric will have an oblate structure, for $\theta=\pi/2$ the potential $\widetilde \Phi \sim \Phi$, but as $\theta =0$ the potential grows as $r^2$. We now check that we have not introduced any singularities to this metric by rotating the original static metric. Possible locations for singularities in this metric occur at $r^{2}+a^{2}\cos^{2}\theta=0$, which would occur if $r=0$ and $\theta=\pi /2$.  We determine the Kretschmann scalar at $r\rightarrow0$ and $\theta=\pi / 2$ to be
\begin{equation}
\begin{aligned}
K=\frac{a^{4}\left(8\pi M_{p}^{2}\right)^{2}}{\left(8\pi M_{p}^{2} + mM \right)^{2}}\,,
\end{aligned}
\end{equation}
which remains finite.

\section{Frame Dragging}

Another interesting property of a rotating metric is the ergosphere, this is a region of the space time where it is impossible for an observer to stand still. In order for an observer to stand still in a region of space their 4-trajectory, $ {\textbf S}$,  must be time like, ${\textbf S}<0$. The 4-trajectory is determined as ${\textbf S}=g_{ab}T^{a}T^{b}$, where $T^{a}=dx^{a}/dt$ is a tangent vector to the worldline of our observer. If the observer is standing still then this vector is $T^{a}=(1,0,0,0)$. Our 4-trajectory is therefore ${\textbf S}=g_{tt}=-(1+2\widetilde{\Phi})$. Since $\widetilde{\Phi} < 1$, but always positive, we will never enter a region of space where ${\textbf S}$ is not time like, and as such our metric contains no ergosphere. Although there is no ergoregion in our metric, we would still expect to see a frame dragging effect, which we can compare to the Kerr metric. An equivalent co-rotating metric would have to have an angular velocity of $\Pi=d\phi/dt=-g_{t\phi}/g_{\phi\phi}$, this angular velocity is the frame dragging effect that an observer would experience at the particular location in the space time. We will begin our analysis in the region of space parametrized by $r> Gm + \sqrt{G^{2}m^{2}-a^{2}\cos^{2}\theta}$. Since in the Kerr metric, $a\leq Gm$, therefore restricting ourselves in the linear regime yields, $a\leq Gm \ll 1/M<r$, which implies that $r^{2}+a^{2}\approx r^{2}$. Working in the equatorial plane, where $\theta=\pi/2$ we can write the frame dragging effect for both Kerr, and for Eq.~(\ref{DNJMetric})
\begin{equation}\label{angvel}
\Pi_{Kerr}\sim \Pi_{IDG}\sim \frac{2amG}{r^{2}\left(r+2mG \right)}.
\end{equation}
Next consider the region of $r<Gm+\sqrt{G^{2}m^{2}-a^{2}\cos^{2}\theta}$, where we expect to see some deviation between the two frame dragging effects. We again work in the equatorial plane. For the Kerr metric the frame dragging effect is written as
\begin{equation}
\Pi_{Kerr}=\frac{2amG}{r\left(r^{2} + a^{2} + \frac{2Gm}{r} a^{2}\right)}
\end{equation}
and the frame dragging effect for the IDG metric, Eq.~(\ref{DNJMetric}) is given as 
\begin{equation}
\Pi_{IDG} = \frac{amMG}{r^{2}+a^{2} + r^{2}mMG + a^{2}mMG}
\end{equation}
In the limit, when $r\rightarrow 0$ we get that the frame dragging effect for both metrics becomes 
\begin{equation}
\Pi_{Kerr}=\frac{1}{a}\,\,\, \text{and}\,\,\, \Pi_{IDG}\approx\frac{mMG}{a}.
\end{equation}
One can also show that the frame dragging effect for the Kerr would always dominate in the region $r<Gm+\sqrt{G^{2}m^{2}-a^{2}\cos^{2}\theta}$, or for $r< 2Gm$.

 Before we conclude, let us consider the results obtained from the Gravity Probe B satellite, see~\cite{Everitt:2011hp}. The satellite contains a set of gyroscopes in low circular polar orbit with altitude $r = 650$ km from the surface of earth. According to GR, the gyroscopes will undergo a geodesic precession in the orbital plane, as well as a Lense-Thirring precession~\cite{Thirring} in the plane of the Earth's equator. The Lense-Thirring precession is related to the off diagonal components of the metric tensor of a rotating gravitational source, so its experimental verification will test the Einstein theory of gravitation. The value for the geodesic precession as predicted by GR is $\Omega_{G(GR)}=6606$  milliarcsec/year,~see~\cite{Adler:1999yt}, and was measured by the Gravity Probe B to be $\Omega_{G}=6602\pm18$ milliarcsec/year~\cite{Everitt:2011hp}. The predicted value for the Lense-Thirring preession is $\Omega_{LT(GR)}=39.2$ milliarcsec/year and was measured as $\Omega_{LT}=37.2\pm 7.2$ milliarcsec/year \cite{Everitt:2011hp}. The Lense-Thirring precession in Cartesian coordinates can be recast as \cite{Adler:1999yt}:
\begin{equation}
\begin{aligned}
\Omega_{G}=& -\frac{3}{2}\nabla\Phi\times\vec{\textbf{V}}\;\;\; \text{and}\;\;\;\Omega_{LT}=\frac{1}{2}\nabla\times \vec{h},
\end{aligned}
\end{equation}
respectively, where $h_{0i}$ are our off diagonal terms and $\vec{\textbf{V}}$ is the four velocity of our orbiting gyroscope. By using the metric that we have obtained in Eq.~(\ref{MetricCart}), and applying the definition of $\omega$, we obtained:
\begin{equation}\label{IDGPress}
\begin{aligned}
\Omega_{G(IDG)}=\left(\textbf{Erf}\left(\frac{Mr}{2}\right)-\frac{Mr}{\sqrt{\pi}}e^{-\frac{M^{2}r^{2}}{4}} \right)\Omega_{G(GR)},\\
\Omega_{LT(IDG)}=\left(\textbf{Erf}\left(\frac{Mr}{2}\right)-\frac{Mr}{\sqrt{\pi}}e^{-\frac{M^{2}r^{2}}{4}} \right)\Omega_{LT(GR)},\\
\end{aligned}
\end{equation}
where we have used the off diagonals as given in Eq. (\ref{OffDiagCart}), and the GR results are given by~\cite{Adler:1999yt}:
 \begin{equation}
\begin{aligned}
\Omega_{G(GR)} = \frac{3GM}{2r^{3}}\left(\vec{r}\times \vec{\textbf{V}} \right),\;\; \Omega_{LT(IDT)}= \frac{2G}{r^{3}}\vec{J}.
\end{aligned}
\end{equation}
Let us constrain the value of $M$, by taking into account when the results from IDG matches that of the GR results, which is well within the errors bars of Gravity probe B. By taking the value of $r=7021$~km, and by approximating the gyroscopes orbit to be nearly perfectly round. Note that Eq.~(\ref{IDGPress}) reduces to that of GR values when $Mr/2\gtrsim 1.5$. Plugging in the value of $r$, we obtain $M\gtrsim 10^{-14}eV$, which is a much weaker constraint than the deviation from the $1/r$-law of gravity obtained in Ref.~\cite{Edholm:2016hbt}.
A similar bound on $M$ can be inferred using the data by the LARES (LAser RElativity Satellite) mission \cite{laresdata} designed to probe the frame-dragging and the Lense-Thirring effect $(0.1-1)\%$ \cite{ciufolini} of the value predicted by GR (LARES's body was inserted in an orbit with 1450~km of perigee, eccentricity $9.54\times10^{-4}$, inclination of $69.5\pm 1$ degrees).

\section{Conclusion}

To conclude we have shown that it is possible to derive a metric for a rotating mass within ghost free and singularity free IDG. Furthermore, in the IR limit we recover exactly the GR metric for a rotating mass. Using the results of Gravity Probe B see are able to place a lower bound on $M\geq 10^{-14}$~eV, although the bound itself is not very impressive. However, improvements of the bounds on $M$ could be obtained by the next generation of space-based tests, such as LARES experiment (see for example \cite{ciufolini}).
Nevertheless, the analysis demonstrates the success of IDG in the IR limit. The non-rotating metric could be further helpful to understand the properties of rotating astrophysical blackholes and primordial blackholes, one of the hot topics of research given the remarkable success of detecting the gravitational waves from LIGO/VIRGO and future gravitational wave observatories. Results of this paper, indeed, are based on linear approximations, but it is expected that the non-locality at a quantum level may prevent the formation of singularity even for astrophysical objects by shifting the scale of non-locality $M$ to the infrared length scales for objects as heavy as LIGO/VIRGO candidates~\cite{Koshelev:2017bxd}. However, a separate investigation is definitely desirable given the importance of experimental data we have in connection with blackhole mergers. Moreover, an inspection of Eq.~(\ref{angvel}) suggests the possibility to have different angular velocities in IDG as compared to the standard Kerr's geometry. This might have potential observable implications in the ring-down phase and the subsequent echoes. All these new possibilities are under investigation.

\section{Acknowledgements}
AM would like to thank the hospitality of Universities of the Witwatersrand and of Cape town, where part of the research was conducted.

\end{document}